\documentclass[a4paper,normalheadings,pointlessnumbers,11pt]{scrartcl}
\pdfoutput=1
\usepackage{graphicx, color}

\usepackage[paper=a4paper,left=30mm,right=30mm,top=30mm,bottom=30mm]{geometry}

\usepackage[square,numbers]{natbib}
\usepackage{lmodern}
\usepackage[utf8]{inputenc}
\usepackage[T1]{fontenc}
\usepackage[english]{babel}
\usepackage{setspace}
\usepackage{graphicx}
\usepackage{caption}
\usepackage{hyperref}
\usepackage{color}
\usepackage{subcaption}
\usepackage{graphicx}
\usepackage{alltt}
\usepackage{enumerate}
\usepackage{booktabs}
\usepackage{amsmath}
\usepackage{amsfonts}
\usepackage{amssymb}
\usepackage{amsthm}
\usepackage{hyperref}
\usepackage{caption}
\usepackage{paralist}
\usepackage{subcaption}
\usepackage{scrpage2}

\usepackage{amsmath}
\usepackage{amsfonts}
\usepackage{amssymb}
\usepackage{amsthm}

\theoremstyle{plain}

\theoremstyle{definition}   %% 

\usepackage{tikz}
\usepackage{algorithm}
\usepackage[noend]{algpseudocode} 
\usepackage{caption}
\algrenewcommand{\algorithmicrequire}{\textbf{Input:}}
\algrenewcommand{\algorithmicensure}{\textbf{Output:}}
\algrenewcommand{\algorithmicforall}{\textbf{for each}}

\theoremstyle{remark}

\theoremstyle{definition}
\newtheorem{defi}{Definition}[section]
\theoremstyle{theorem}

\newcommand{\B}{\mathcal{B}}

\newcommand{\name}[1]{{\texttt{#1}}}

\title{Automated Cryptanalysis of Bloom Filter Encryptions of Health Records}
\author{Martin Kroll \quad \quad \quad Simone Steinmetzer\\
\vspace{-5mm}\\
\textit{University of Duisburg-Essen}\\
\small \texttt{\{martin.kroll, simone.steinmetzer\}@uni-due.de}
}

\date{\today}

\begin{document}

\definecolor{DodgerBlue4}{RGB}{016,078,139}

\addtokomafont{title}{\color{DodgerBlue4}}
\addtokomafont{section}{\color{DodgerBlue4}}

\maketitle

\abstract{
Privacy-preserving record linkage with Bloom filters has become increasingly popular in medical applications, since Bloom filters allow for probabilistic linkage of sensitive personal data.
However, since evidence indicates that Bloom filters lack sufficiently high security where strong security guarantees are required, several suggestions for their improvement have been made in literature.
One of those improvements proposes the storage of several identifiers in one single Bloom filter.
In this paper we present an automated cryptanalysis of this Bloom filter variant.
The three steps of this procedure constitute our main contributions:
(1) a new method for the detection of Bloom filter encrytions of bigrams (so-called \emph{atoms}), (2) the use of an optimization algorithm for the assignment of atoms to bigrams, (3) the reconstruction of the original attribute values by linkage against bigram sets obtained from lists of frequent attribute values in the underlying population.
To sum up, our attack provides the first convincing attack on Bloom filter encryptions of records built from more than one identifier.
}


\section{Introduction}\label{sec:introduction}

\noindent Record linkage between databases containing information on individual people is popular in a large number of medical applications, for example
the identification of patient deaths~\cite{Jones2005}, the evaluation of disease treatment~\cite{Newman1997} and the linkage of cancer registries in epidemiology~\cite{Van1990}.
In many applications data sets are merged using personal identifiers such as forenames, surnames, place and date of birth.
Due to privacy concerns this has to be done via privacy-preserving record linkage (PPRL).
However, since personal identifiers often contain typing or spelling errors, encrypting the identifier values and linking only those that match exactly does not provide satisfactory results.
Therefore, to allow for errors in encrypted personal identifiers, in many European countries encrypted phonetic codes, such as Soundex codes, are commonly used, especially by cancer registries.
As the performance of these codes is still non satisfactory, several novel privacy-preserving record linkage methods have been suggested during the last years.
For example Schnell et al. \cite{Schnell2009} developed a method based on Bloom filters.
Bloom-filter-based record linkage has already been used in medical applications in a number of different countries \cite{Kuhni2011,NapoleaoRocha2013,Randall2014,Schnell2014}.

Another frequently applied privacy-preserving record linkage method uses anonymous linking codes \cite{Herzog2007}.
The basic principle of an anonymous linking code is to standardize all particular identifiers of a record (removal of certain characters and diacritics, use of upper case letters), to concatenate them to a single string and finally to put this single string into a cryptographic hash function.
By combining this principle with Bloom filters, Schnell et al. \cite{Schnell2011} first developed a novel error-tolerant anonymous linking code, called Cryptographic Longterm Key (CLK).
Instead of encrypting every single identifier from a record of several identifiers through a Bloom filter, multiple identifiers are stored in one single Bloom filter, called CLK.
Tests on several databases showed that CLKs yield good linkage properties, superior to well-known anonymous linking codes \cite{Schnell2011}.

Only recently Randall et al. \cite{Randall2014} presented a study on 26 million records of hospital admissions data and showed that privacy-preserving record linkage with Bloom filters built from multiple identifiers is applicable to large real-world databases without loss in linkage quality.

However, only little research on the security of Bloom filters built from more than one identifier has yet been published (see subsection~\ref{extent_research}).
In several countries, this lack of research prevents the widespread use of Bloom filter encryptions for real-world medical databases (such as cancer registries) where the anonymity of the individuals has to be guaranteed.
For example, in its \emph{Beyond 2011 Programme} the British Office for National Statistics investigated several methods for linking sensitive data sets~\cite{Office2013}.
The investigators came to the conclusion that none of the ``(...) recent innovations, such as bloom filter encryption (...)" can be recommended because they  ``(...) have not been fully explored from an accreditation perspective".
Thus, research showing drawbacks of the recent Bloom filter techniques is important because it guides the direction for future research and might motivate further development of the recent procedures.
In this paper, we intend to investigate this issue in detail by giving the first convincing cryptanalysis of Bloom filter encryptions built from more than one identifier.

\section{Background}

\noindent In 1970, Burton Howard Bloom \cite{Bloom1970} introduced a novel approach that permits
the efficient testing of set membership through a probabilistic space-efficient data structure.
A \emph{Bloom filter} is a bit array of length $L$, which at first contains zeros only.
Let $S \subseteq \mathcal{U}$ be a subset of a universe $\mathcal{U}$. Then $S$
can be stored in a Bloom filter $\B = \B(S) = (b_{0} , \dots , b_{L-1})$ in the following way:
Each element $s \in S$ is mapped via $k$ different hash functions 
$h_{0}, \dots, h_{k-1}: S \longrightarrow \{ 0, \dots, L-1 \}$ and all the corresponding bit positions $b_{h_{0}(s)}, \dots, b_{h_{k-1}(s)}$ are set to one.
Once a bit position is set to one this value no longer changes.

Furthermore, to test whether an item $u \in \mathcal{U}$ from the universe is contained in  $S$, $u$ is hashed through the $k$ hash functions $h_{0}, \dots, h_{k-1}$ as well.
Consequently, if all bit positions $b_{h_{0}(u)}, \dots, b_{h_{k-1}(u)}$ in the Bloom filter are set to one, then $u \in S$ holds with high probability.
However, false positive values can occur when the ones on positions $h_{0}(u), \dots, h_{k-1}(u)$ are caused by two or more different elements $u$.
Then the test indicates $u \in S$ although this is not the case. 
Otherwise, if at least one bit position in the two Bloom filters varies, $u$ clearly is no member of $S$.

\subsection{PPRL with Bloom Filters Built from Multiple Identifiers}

In~\cite{Schnell2009} Bloom filters were used in privacy-preserving record linkage for the first time.
This approach was expanded to Cryptographic Longterm Keys in \cite{Schnell2011}.

In common PPRL protocols two data owners A and B agree on a set of identifiers that occur in both of their databases. 
Next, these identifiers are standardized, then padded with blanks at the beginning and the end, and finally split into substrings of two characters. 
Each substring of the first identifier corresponding to a record is mapped to the first Bloom filter via several hash functions.
Afterwards, each substring of the second identifier, corresponding to the same record, is mapped through another set of hash functions to the first Bloom filter as well.
This procedure is repeated until all identifiers of the first record are stored in the first Bloom filter.
Next, all identifiers corresponding to the second record of the database are mapped through the utilized hash functions to a second Bloom filter and so on.
Performing this procedure for all entries of the database results in a set of Bloom filters where each Bloom filter is built from multiple identifiers.
Thus, the similarity of the Bloom filters is a measure for the similarity of the encoded identifiers.
Usually, the linkage of the two databases is conducted by a third party C.

Because of the specific structure of Bloom filters, record linkage based on Bloom filters built from multiple identifiers allows for errors in the encrypted data.
Therefore, they can be applied to linking large data sets such as national medical databases~\cite{Randall2014}.

\subsection{Extant Research: Attacks on Bloom Filters of One or More Identifiers}\label{extent_research}

To the best of our knowledge, only two ways of attacking Bloom filters of one identifier and one way of attacking Bloom filters of multiple identifiers are known so far.

The first cryptanalysis of Bloom filters was published in 2011.
Kuzu et al. \cite{Kuzu2011} sampled 20,000 records from a voter registration list and encrypted the substrings of two characters from the forenames through 15 hash functions and Bloom filters of length 500 bits.
Their attack consisted in solving a constraint satisfaction problem (CSP).
Through a frequency analysis of the fornames and the Bloom filters and by applying their CSP solver to the problem, Kuzu et al. were able to decipher approximately 11\% of the data. 

In contrast, Niedermeyer et al. \cite{Niedermeyer2014} proposed an attack on 10,000 Bloom filters built from encrypted German surnames that were considered to be a random sample of a known population.
For the generation of the Bloom filters 15 hash functions and Bloom filter length 1,000 were used.
Then they conducted a manual attack based on the frequencies of the substrings of length two, which they derived from the German surnames.
Thus, Niedermeyer et al. deciphered the 934 most frequent surnames of 7,580 different ones, which corresponds to approximately 12\% of the data set.
However, their attack is not limited to the most frequent names and could be extended to the decipherment of nearly all names. 

In 2012 Kuzu et al. \cite{Kuzu2012} showed an attack on Bloom filters built from multiple identifiers.
They applied their constraint solver to forename and surname, as well as forename, surname, city and ZIP code, of 50,000 randomly selected records from the North Carolina voter registration list. 
However, they were not able to mount a successful attack.
Thus, Kuzu et al. supposed that combining multiple personal identifiers into a single Bloom filter would offer a protection mechanism against frequency attacks.
Although they suspected that their attack did not uncover all vulnerabilities of the Bloom filter encodings, they showed that the CSP for multiple identifiers is intractable to solve by their constraint solver.

\subsection{Our Contribution}

In this paper we present a fully automated attack on a database containing forenames, surnames and the relevant place of birth as well.
All records are considered to be a random sample of a known population.
We suppose that the attacker only knows some publicly available lists of the most common forenames, surnames and locations.
The attack is based on analyzing the frequencies and the combined occurence of substrings of length two from the identifiers of these lists.
Furthermore, we are interested in recovering as many identifiers as possible.
Our cryptanalysis was implemented using the programming languages Python and C++.

%-------- Section ------------

\section{Encryption}\label{Encryption}

\noindent In this section some basic notation is introduced and the encrypting procedure is described.

In record linkage scenarios, strings are usually standardized through transformations such as capitalization of characters or removal of diacritics~\cite{Randall2013}.
After this \emph{preprocessing step} all strings contain only tokens from some predefined alphabet $\Sigma$.
Throughout this article, we use the canonical alphabet $\Sigma := \{\name{A}, \name{B}, \dots ,\name{Z}, \name{\textvisiblespace}\}$, where $\text{\textvisiblespace}$ denotes the padding blank.
Thus, for example the popular German surname \name{M\"uller} is transformed to \name{\textvisiblespace MUELLER\textvisiblespace} in the preprocessing step.
As usual, we denote substrings of two characters with \emph{bigrams} and the set containing all the bigrams with $\Sigma^2$, i.e.\\
$\Sigma^2=\{\name{\textvisiblespace \textvisiblespace}, \name{\textvisiblespace A}, \dots, \name{\textvisiblespace Z}, \name{A\textvisiblespace}, \dots, \name{Z\textvisiblespace}, \name{AA}, \dots, \name{ZZ}\}$.

The Bloom filter encryption of a record from a database is created by storing the bigram set associated with this record into a Bloom filter.
The bigram set associated with a record is defined as the set containing the bigrams from all the identifiers.
Here, a distinction between the bigrams occuring in different identifiers is made.
Thus, if the set of identifiers is denoted with $\mathcal I$, the bigram set of a record is a subset of $\mathcal I \times \Sigma^2$.

For example, if we have $\mathcal I=\{\text{surname}, \text{forename}\}$ and the database contains a record, \texttt{Peter M\"uller}, the bigram set $\mathcal S_{\text{record}}$ associated with this record would contain the bigrams $\name{\textvisiblespace P}_f$, $\name{PE}_{f}$, $\name{ET}_f$, $\name{TE}_f$, $\name{ER}_f$, $\name{R\textvisiblespace}_f$, $\name{\textvisiblespace M}_s$, $\name{MU}_s$, $\name{UE}_s$, $\name{EL}_s$, $\name{LL}_s$, $\name{LE}_s$, $\name{ER}_s$ and $\name{R\textvisiblespace}_s$ (the subscript $f$ indicates the bigrams occuring in the forename identifier, the subscript $s$ the ones occuring in the surname identifier).

Next, this bigram set is stored into a Bloom filter $(b_0,\ldots,b_{L-1})$ of length $L$ by means of $k$ independent hash functions
\begin{equation*}
h_i: \mathcal I \times \Sigma^2 \to \{0,\ldots,L-1\}
\end{equation*}
for $i=0,\ldots,k-1$.
In practice, one could alternatively use different hash functions $h_i: \Sigma^2 \to \{0,\ldots,L-1\}$ for the distinct identifiers in order to guarantee that the hash values for distinct identifiers are not the same.

Further, as in~\cite{Niedermeyer2014} we introduce the term \emph{atom} for the specific Bloom filters which occur as the fundamental building blocks of the encryption method.
\begin{defi}[Atom]
Let $L, k \in \mathbb N$ and some hash functions $h_{0},\ldots,h_{k-1}$ be defined as above. Then, a Bloom filter
\begin{equation*}
		\B := (b_{0}, \dots, b_{L-1}) \in \{0,1\}^{L} 
\end{equation*}
is termed an \emph{atom} if there exists a bigram $\beta \in \mathcal I \times \Sigma^2$ such that $b_j=1 \Leftrightarrow h_i(\beta)=j$ for some $i=0,\ldots,k-1$. Such a Bloom filter is called the \emph{atom realized by the bigram $\beta$} and denoted with $\B(\beta)$.
\end{defi}
\noindent

Thus, atoms are special Bloom filters.
Since each bigram is hashed via each $h_i$ for $i=0,\ldots,k-1$, at most $k$ positions in an atom can be set to one.

By combining the atoms of the underlying bigram set of a record with the bitwise OR operation, the Bloom filter of a record is composed as
\begin{equation*}
	\B(\text{record}) = \underset{\beta \in \mathcal S_{\text{record}}}{\bigvee} \B(\beta),
\end{equation*}
where $\bigvee$ denotes the bitwise OR operator.

\begin{figure*}[b!t!]
	\begin{center}
    \begin{tikzpicture}
	\draw[step=0.5cm](0,0)grid(3,0.5);
	\draw[](3,0.5)--(7,0.5);
	\draw[](3,0)--(7,0);
	\draw[step=0.5cm](6.5,0)grid(10,0.5);
	\node at (5,2) {$\name{ER}_f$};
	\draw [thick,->] (5,1.75)--node[anchor=south east]{$h_{1}$}(0.75,0.5) ;
	\draw [thick,->] (5,1.75)--node[anchor=south west]{$h_{0}$}(8.75,0.5) ;
	\draw [thick,->] (5,1.75)--node[anchor=north west]{$h_{2}$}(2.75,0.5) ;
	\node at (5,1) {\dots};
	\node at (-0.5,0.25) {0};
	\node at (0.25,0.25) {0};
	\node at (0.75,0.25) {1};
	\node at (1.25,0.25) {0};
	\node at (1.75,0.25) {0};
	\node at (2.25,0.25) {0};
	\node at (2.75,0.25) {1};
	\node at (3.25,0.25) {0};
 	\draw[](3.5,0)--(3.5,0.5);
 	\node at (5,0.25) {\dots};
 	\draw[](6.5,0)--(6.5,0.5);
 	\node at (6.75,0.25) {0};
 	\node at (7.25,0.25) {0};
 	\node at (7.75,0.25) {0};
 	\node at (8.25,0.25) {0};
 	\node at (8.75,0.25) {1};
 	\node at (9.25,0.25) {0};
 	\node at (9.75,0.25) {0};
	\node at (10.5,0.25) {999};
	\end{tikzpicture}
 	\end{center}
    \begin{center}
	\begin{tikzpicture}
	\node at (5,-2) {$\name{ER}_s$};
    \node at (5,-1) {\dots};
	\draw [thick,->] (5,-1.75)--node[anchor=north east]{$h_{2}$}(0.25,0) ;
	\draw [thick,->] (5,-1.75)--node[anchor=west]{$h_{3}$}(1.75,0) ;
	\draw [thick,->] (5,-1.75)--node[anchor=south west]{$h_{4}$}(3.25,0) ;
    \draw [thick,->] (5,-1.75)--node[anchor=south east]{$h_{0}$}(7.25,0) ;
    \draw [thick,->] (5,-1.75)--node[anchor=north west]{$h_{1}$}(8.75,0) ;
    \draw[step=0.5cm](0,0)grid(3,0.5);
	\draw[](3,0.5)--(7,0.5);
	\draw[](3,0)--(7,0);
	\draw[step=0.5cm](6.5,0)grid(10,0.5);
	\node at (-0.5,0.25) {0};
	\node at (0.25,0.25) {1};
	\node at (0.75,0.25) {0};
	\node at (1.25,0.25) {0};
	\node at (1.75,0.25) {1};
	\node at (2.25,0.25) {0};
	\node at (2.75,0.25) {0};
	\node at (3.25,0.25) {1};
 	\draw[](3.5,0)--(3.5,0.5);
    \node at (5,0.25) {\dots};
 	\draw[](6.5,0)--(6.5,0.5);
 	\node at (6.75,0.25) {0};
 	\node at (7.25,0.25) {1};
 	\node at (7.75,0.25) {0};
 	\node at (8.25,0.25) {0};
 	\node at (8.75,0.25) {1};
 	\node at (9.25,0.25) {0};
 	\node at (9.75,0.25) {0};
	\node at (10.5,0.25) {999};
	\end{tikzpicture}
	\end{center}
    \centering
	\caption[BF]{Two different atoms of the bigram \name{ER}. These atoms are realized when instances of \name{ER} occur in distinct identifiers.}
	\label{fig:atoms}
\end{figure*}
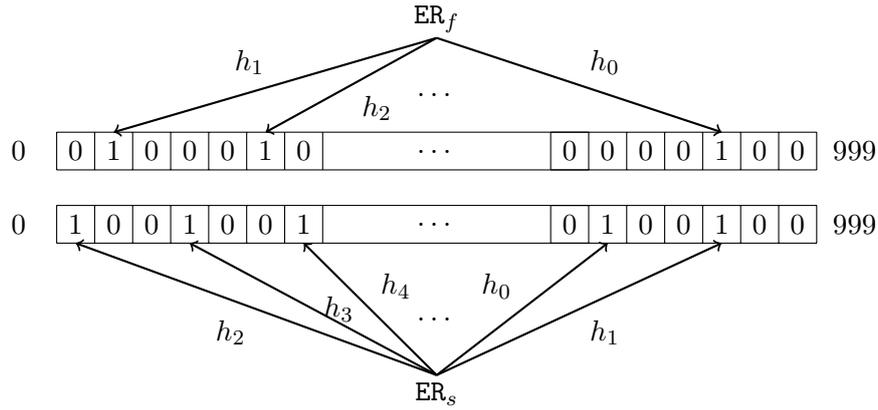

Note that the same bigram from $\Sigma^2$ is hashed differently if it occurs in distinct identifiers.
This is illustrated in Figure~\ref{fig:atoms} for the example of the bigram $\name{ER}$ which occurs in the record \name{Peter M\"uller} both in the surname and the forename identifier.

Mapping each bigram of the forename \name{Peter} with $k$ hash functions results in six atoms;
for the surname \name{M\"uller}, we get eight atoms.
Thus, the separate Bloom filters for these identifiers might be composed as illustrated in Figure~\ref{fig:bspbf}. 
\begin{figure*}[t!b!]
	\begin{eqnarray*}
% (998+6*0:14)%%1000 998   4  10  16  22  28  34  40  46  52  58  64  70  76  82
% (991+6*0:14)%%1000 991 997   3   9  15  21  27  33  39  45  51  57  63  69  75
% (993+2*0:14)%%1000 993 995 997 999   1   3   5   7   9  11  13  15  17  19  21
% (993+5*0:14)%%1000 993 998   3   8  13  18  23  28  33  38  43  48  53  58  63
% (997+4*0:14)%%1000 997   1   5   9  13  17  21  25  29  33  37  41  45  49  53
% (999+2*0:14)%%1000 999   1   3   5   7   9  11  13  15  17  19  21  23  25  27
				& 0000100000 \dots 0000000010& \quad \B(\name{\textvisiblespace P}_f)\\[-1ex] %998+6j
		\vee	& 0001000001 \dots 0100000100& \quad \B(\name{PE}_f)\\[-1ex] %991+6j
		\vee	& 0101010101 \dots 0001010101& \quad \B(\name{ET}_f)\\[-1ex] %993+2j
		\vee	& 0001000010 \dots 0001000010& \quad \B(\name{TE}_f)\\[-1ex] %993+5j
		\vee	& {\color{black}{0100010001 \dots 0000000100}}& \quad \B(\name{\color{red}ER}_f)\\[-1ex] %997+4j
		\vee	& 0101010101 \dots 0000000001& \quad \B(\name{R\textvisiblespace}_f)\\[-2.5ex] %999+2j
				&\begin{tikzpicture}
					\draw[](2,0)--(8,0);
				 \end{tikzpicture}\\[-0.5ex]
				& 0101110111 \dots 0101010111& \quad \B(\name{Peter})
\end{eqnarray*}

\begin{eqnarray*}
				& 0000000100 \dots 0000000001& \quad \B(\name{\textvisiblespace M}_s)\\[-1ex] %999+8j
		\vee	& 0010000000 \dots 0100000000& \quad \B(\name{MU}_s)\\[-1ex] %991+11j
		\vee	& 0000100000 \dots 0010000010& \quad \B(\name{UE}_s)\\[-1ex] %992+6j
		\vee	& 1000000010 \dots 0010000000& \quad \B(\name{EL}_s)\\[-1ex] %992+8j
		\vee	& 0100001000 \dots 0100001000& \quad \B(\name{LL}_s)\\[-1ex] %991+5j
		\vee	& 1000000100 \dots 0001000000& \quad \B(\name{LE}_s)\\[-1ex] %993+7j
		\vee	& {\color{black}{1001001001 \dots 0000100100}}& \quad \B(\name{\color{red}ER}_s)\\[-1ex] %994+3j
		\vee	& 0010001000 \dots 0000000010& \quad \B(\name{R\textvisiblespace}_s)\\[-2.5ex] %998+4j
				&\begin{tikzpicture}
					\draw[](2,0)--(8,0);
				 \end{tikzpicture}\\[-0.5ex]
				& 1111101111 \dots 0111101111& \quad \B(\name{M\"uller})
	\end{eqnarray*}
    \centering
	\caption[OR]{Bloom filters of the forename \name{Peter} and the surname \name{M\"uller}, composed of the atoms belonging to the underlying bigrams.}
	\label{fig:bspbf}
\end{figure*}

The final Bloom filter for the record \name{Peter M\"uller} is composed by appling the bitwise OR operation to the separate Bloom filter encryptions of the distinct identifiers.
This is demonstrated in Figure~\ref{fig:fullname}.

\begin{figure*}[t!b!]
	\begin{eqnarray*}
				& 0101110111 \dots 0101010111& \quad \B(\name{Peter})\\[-1ex]
		\vee	& 1111101111 \dots 0111101111& \quad \B(\name{M\"uller})\\[-2.5ex]
        		&\begin{tikzpicture}
					\draw[](2,0)--(8,0);
				 \end{tikzpicture}\\[-0.5ex]
                 & 1111111111 \dots 0111111111& \quad \B(\name{entire record})
	\end{eqnarray*}
	\centering
	\caption[CLK]{The Bloom filter of the record \name{Peter M\"uller} is obtained by applying the bitwise OR operation to the Bloom filter encryptions of the separate identifiers.}
    \label{fig:fullname}
\end{figure*}

In practice, the Bloom filter encryption of a record might contain a mixture of string valued identifiers (such as forename, surname or place of birth) and also numerical identifiers, such as date of birth.
However, in this paper we restrict ourselves to the case of string valued attributes only, albeit our cryptanalysis proposed below is not limited to such attributes.

\subsection*{Assumptions}

In many record linkage scenarios, it is supposed that a semi-trusted third party conducts the record linkage between two encrypted databases.
In this paper we assume a data set containing Bloom filters built from multiple identifiers that is sent to a semi-trusted third party.
This third party acts as the adversary and tries to infer as much information as possible from the record encryptions.
We further suppose that the attacker has knowledge of the encryption process.

For our scenario we generated 100,000 Bloom filters built from standardized German forenames, surnames and cities according to the distribution in the population.
The identifiers were truncated after the tenth letter, padded with blanks, respectively, and were broken into bigrams.
Then the bigrams were hashed through $k=20$ hash functions into Bloom filters of length $L=1,000$.
As proposed in~\cite{Schnell2009} and~\cite{Schnell2011}, we used the so-called \emph{double hashing scheme} for the generation of $k$ hash functions from two hash functions $f$ and $g$. This double hashing scheme is defined via the equation
\begin{equation}\label{eqhash}
h_i = (f + i \cdot g) \bmod L \quad \text{ for } i=0,\ldots,k-1
\end{equation}
\noindent
and was originally proposed in~\cite{Kirsch2008} as a simple hashing method for Bloom filters yielding satisfactory performance results.

In our cryptanalysis we assume that the adversary knows that the hash values are generated in accordance with equation~\eqref{eqhash}.
It is self-evident that s/he must not have direct access to the hash functions $f$ and $g$ since this would permit the adversary to check whether a specific bigram is contained in a given Bloom filter.

Note that the double hashing scheme has also been used for the generation of Bloom filters by Kuzu et al.~\cite{Kuzu2012}.
However, in that paper the knowledge of the \emph{double hashing scheme} was not exploited in their cryptanalysis. 
  
%-------- Section ------------

\section{Cryptanalysis}\label{CryptanaCLK}

\noindent
This section provides a detailed description of the deciphering process.
At first we try to detect the atoms that are contained in the given Bloom filters.
Then, we assign bigrams to these atoms by means of an optimization algorithm.
Finally, the original attributes are reconstructed from the atoms.

Our approach for the development of a fully automated attack
is based on previous results on the automated cryptanalysis of simple substitution ciphers presented by Jakobsen \cite{Jakobsen1995}.
We give a short account of Jakobsen's results in order to motivate our procedure.

\subsection{Automated Cryptanalysis of Simple Substitution Ciphers}

The encryption of a plaintext message through a \emph{simple substitution cipher} is defined by a permutation of the underlying alphabet $\Sigma$.
For instance, the message \texttt{HELLO}\textvisiblespace\texttt{LISBON} with tokens from the alphabet $\Sigma = \{ \texttt{\textvisiblespace}, \texttt{A}, \texttt{B}, \dots, \texttt{Z}\}$ could be encrypted as \texttt{RVUUYJUOWAYL}.

\noindent It is well known that this kind of encryption can be broken easily by means of a frequency analysis. 
However, just replacing the i-th frequent character in the ciphertext with the i-th frequent character in the underlying language will usually not lead to the correct decipherment (even for longer messages).
This is commonly compensated for by taking bigram frequencies into consideration as well.

The expected bigram frequencies can be obtained from a
training data set composed of the underlying language and stored in a quadratic matrix $E$ (in the above example a $27 \times 27$ matrix), where the entry $e_{ij}$ is equal to the relative proportion of the bigram $c_{i}c_{j}$ in the training text corpus and $c_{i}$ denotes the i-th character of the alphabet.
Analogously, the bigram frequencies of the ciphertext can be stored in a matrix $D$.

The algorithm proposed by Jakobsen \cite{Jakobsen1995} was intended to find a permutation $\sigma_{\text{opt}}$ of the alphabet such that the objective function $f$ defined via
\begin{equation}
	f(\sigma) := \sum_{i,j} \lvert d_{\sigma(i)\sigma(j)} - e_{ij} \rvert
\end{equation}
\noindent
was minimized. 
The algorithm starts with the initial permutation that reflects the best assignment between single characters in the plaintext and the ciphertext with respect to their relative frequency.
In each step of the algorithm two elements of the currently best permutation $\sigma_{\text{opt}}$ are swapped, leading to a new candidate permutation $\sigma$.
If $f(\sigma) < f(\sigma_{\text{opt}})$ holds, the current permutation is updated to $\sigma$, otherwise $\sigma$ is discarded and a new candidate $\sigma$ is generated by swapping two other elements of $\sigma_{\text{opt}}$.
This is repeated until no swap leads to a further improvement of the objective function $f$.
Throughout this paper we use the same strategy as Jakobsen in~\cite{Jakobsen1995}, in order to determine the elements of the current
permutation to be swapped. 
For a more detailed description of Jakobsen's method in the case of simple substitution ciphers we refer the reader to the original paper~\cite{Jakobsen1995}.
Figure 2 in~\cite{Jakobsen1995} shows that a ciphertext of length 600 built by a simple substitution cipher can be entirely broken by this method.
It is clear that some modification of Jakobsen's original algorithm is necessary in order to make it applicable in our setting as well.
In particular, the definitions of the matrices $D$ and $E$ must be changed.
Their adopted definitions are introduced in subsection~\ref{subsec:correlation}.

\subsection{Atom Detection}

As in \cite{Niedermeyer2014}, the basic principle of our approach consists in the detection of atoms, which represent the encryption of one single bigram only.
Since the Bloom filter of a string is created by the superposition of at least a few atoms, the reconstruction of the atoms given only a set of Bloom filters turns out to be difficult.
Note that this task cannot be solved in a satisfactory manner if Bloom filters are considered isolatedly or in small groups because in this case too many binary vectors will be wrongly classified as atoms.

Let us give a short motivation for our novel method aiming at atom detection.
If the bitwise AND operation is applied to a set of Bloom filters that have one bigram $\beta$ in common, at least all positions set to one by $\beta$ are equal to one in the result.
However, for prevalent bigrams it should be expected that all the other positions are set to zero if a sufficient number of Bloom filters are considered, i.e., the result would be exactly the atom induced by the bigram $\beta$.

Of course, if an adversary has access to a set of Bloom filters, s/he does not a priori know which Bloom filters have a bigram in common.
This obstacle can be avoided as follows:
Under the assumption that the double hashing scheme is being used, the adversary is able to determine for each combination of bit positions from equation~\eqref{eqhash} the set of Bloom filters for which all these positions are set to $1$.
Then, the bitwise AND operation is applied to the set of these Bloom filters.
If the result coincides with the atom, it is considered to be the 
realization of a bigram by the adversary.

The resulting set of atoms was further reduced by discarding atoms of Hamming weight $\sum _{i=0}^{999} b_i$ equal to $1$, $2$, $4$ or $5$ and keeping only atoms of Hamming weight equal to $8$, $10$ or $20$.

Otherwise, too many binary vectors would have been classified incorrectly as atoms.
The probability that an atom has Hamming weight less than $8$ in our setting is equal to $0.008$.
This value can be derived in analogy to Lemma A.1 and the subsequent example in~\cite{Niedermeyer2014}.
We denote the number of atoms found by $n$.
For our specific data set we got $n=$ 1,776. % das ist schon der neue Wert
This result seems reasonable, because the total number of possible atoms is bounded from above by 2,187 and obviously not all of these atoms, in particular atoms realized by rare bigrams, occur in our simulated data.
For each atom $\alpha$ we determined the set of Bloom filters containing this atom, i.e. Bloom filters for which all bit positions of the atom are set to $1$.
We denote the atoms with $\alpha_{1}, \dots, \alpha_{1776}$ according to decreasing frequency.

In the subsequent section we explain how correlations between the occurences of atoms in the Bloom filters and bigrams in a training data set can be used to give adequate definitions of the matrices $D$ and $E$ that serve as the input of Jakobsen's algorithm.

\subsection{Correlation of Atoms and Bigrams}\label{subsec:correlation}

\begin{figure}[]
\centering
\includegraphics[width=0.6\linewidth]{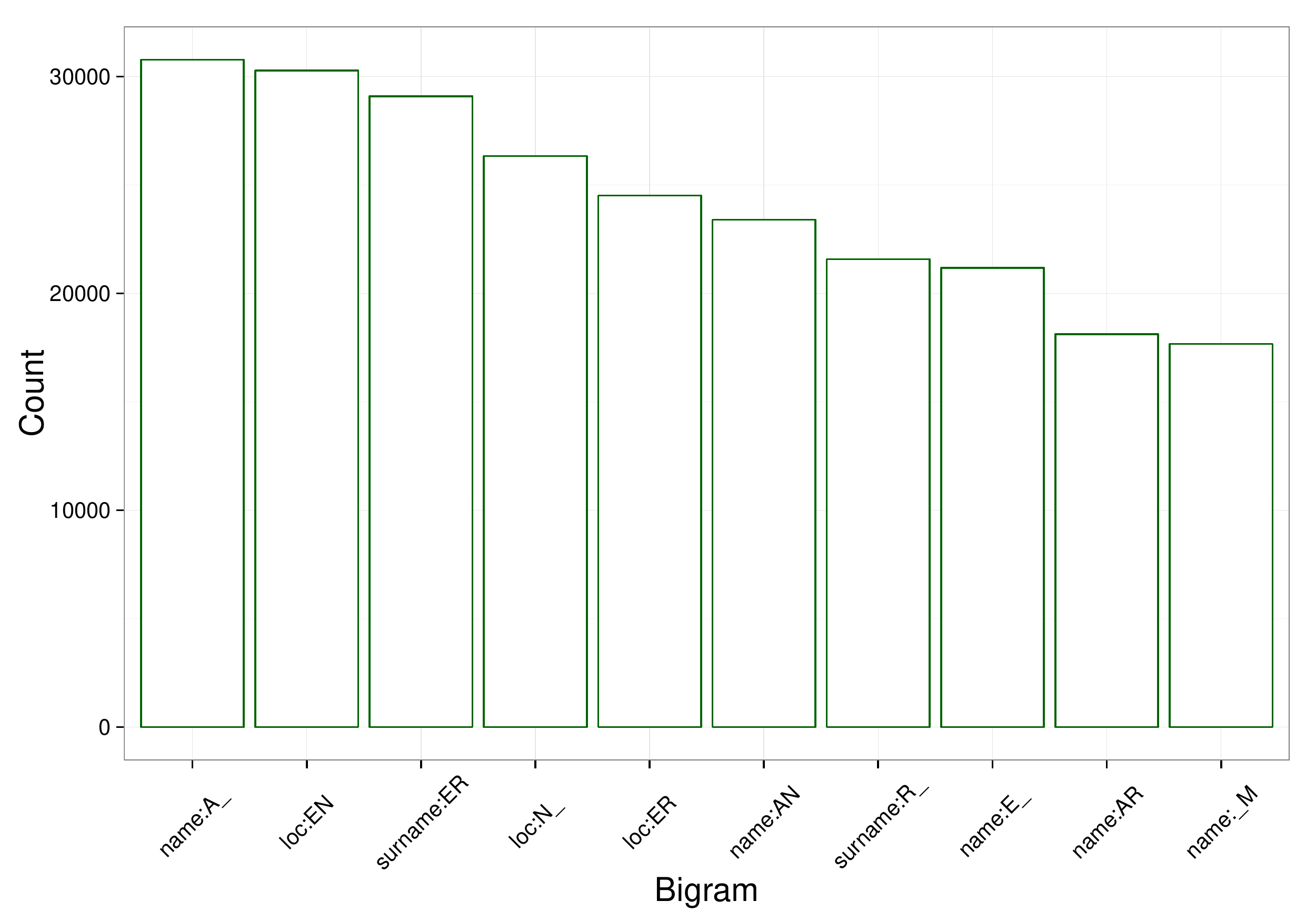}
\caption{Absolute frequencies of the $10$ most frequent bigrams in our training data set.}
\label{fig:bigram_stat}
\end{figure}

A naive assignment of bigrams to atoms is possible only for few frequent bigrams.
For example, if German surnames, given names and birth locations are considered together, the most frequent bigram is $\name{A\textvisiblespace}_f$
(the bigram \name{A\textvisiblespace} in the forename identifier) such that the most frequent atom is likely to be the encryption of this bigram.
The absolute frequencies of the $10$ most frequent bigrams in the considered training data are illustrated in Figure~\ref{fig:bigram_stat}.

Exept for the first few bigrams, the bigram frequencies are too close together such that naive matching is not promising for automatic decipherment.
For this reason, we also took correlations between bigrams into account.
For example, for records sampled from the population of Germany the appearance of the bigram $\name{CH}_s$ in a record makes the appearance of the bigram $\name{SC}_s$ in the same record more likely because the trigram \name{SCH} frequently appears in German surnames.\\
\indent We model this kind of information on the correlation of atoms and bigrams by means of two matrices $D$ and $E$.
Assume that the attribution values of the records built from tokens of the alphabet
$\Sigma = \{\name{\textvisiblespace}, \name{A}, \name{B}, \dots, \name{Z}\}$ are to be encrypted.
Thus, for each (string valued) identifier we have $729$ possible bigrams.
Since the same bigram is encrypted differently for each identifier we have to distinguish between different instances of the same bigram.
In our setting we denote the bigram $\beta$ for the surname, forename and location identifier with $\beta_s$, $\beta_f$ and $\beta_l$, respectively.
Altogether, the set $\Sigma^{2}$ containing all possible bigrams consists of $3 \cdot 729=$ 2,187 elements.\\
\indent Let us now introduce the matrix $E$ containing information about the expected bigram correlations obtained from the training data set.
Note that the training data should be as similar to the encrypted data as possible, e.g. a random sample from the same underlying population as the encrypted data.
If the prevailing Bloom filters are known to contain encryptions of records from the German population, an attacker would try to get access to a comparable database containing the same identifiers.
The attribute values of this training data set are preprocessed analogously to the preprocessing routine before the encryption process.
Then, the bigram sets for all the attribute values are created.
We denote the bigrams with $\beta_{1}, \dots, \beta_{2187}$ according to decreasing frequency.
Let $T$ be the total number of records in the training data set and $t_{ij}$ the number of records that contain both bigram $\beta_{i}$ and bigram $\beta_{j}$.
Then the matrix $E = (e_{ij})_{i,j = 1,\dots, 2187}$ is defined via

$$e_{ij} = \begin{cases} t_{ij}/T &\mbox{if } i\neq j, \\
0 & \mbox{if } i=j. \end{cases}$$

The matrix $D$ is formed in a similar way on the basis of joint appearances of atoms in the Bloom filters.
Let $N$ be the number of Bloom filters for which atoms have been extracted.
We denote the number of Bloom filters that contain both atom $\alpha_{i}$ and atom $\alpha_{j}$ by $b_{ij}$. 
The matrix $D = (d_{ij})_{i,j = 1,\dots, 2187}$ is defined through 
\noindent
$$d_{ij} = \begin{cases} b_{ij}/N &\mbox{if } i\neq j \text{ and } i,j \leq 1776, \\
0 & \mbox{if } i=j \text{ or } \max(i,j)>1776. \end{cases}$$

The procedure suggested by Jakobsen which was described above can now directly be applied to the matrices $D$ and $E$. The pseudocode for the overall algorithm can be found in algorithm~\ref{alg:jakobsen}.

\begin{algorithm}[htbp]
    \caption{\textsc{Optimization algorithm}}
    \label{alg:jakobsen}
	\footnotesize
    \begin{algorithmic}[1]
	\Require $D, E$ as defined in section~\ref{subsec:correlation}
	\Ensure $\sigma_{\text{opt}} \in S_{2187}$ minimizing $$f(\sigma)=\sum \limits_{i,j} \vert d_{\sigma(i)\sigma(j)}-e_{i}e_{j} \vert$$
	\Statex
	\State $\sigma_{\text{opt}}(i)=i \ \forall i$ \Comment Initialization
    \State $\min \gets f(\sigma_{\text{opt}})$
    \State $a, b \gets 1$
    \Repeat
    \State $\sigma \gets \sigma_{\text{opt}}$
    \State $a \gets a+1$
    \If{$a+b \leq 2187$}
    \State $\sigma(a) \gets \sigma_{\text{opt}}(b)$,  $\sigma(b) \gets \sigma_{\text{opt}}(a)$
    \Else
    \State $a \gets 1$, $b \gets b +1$
    \EndIf
    \If{$f(\sigma)<f(\sigma_{\text{opt}})$} \Comment Update
    \State $\min \gets f(\sigma)$
    \State $\sigma_{\text{opt}} \gets \sigma$
    \State $a,b \gets 1$
    \EndIf
    \Until{$b = 2187$}
    \end{algorithmic}
    \end{algorithm}

\begin{figure}[b]
\centering
\includegraphics[width=0.6\linewidth]{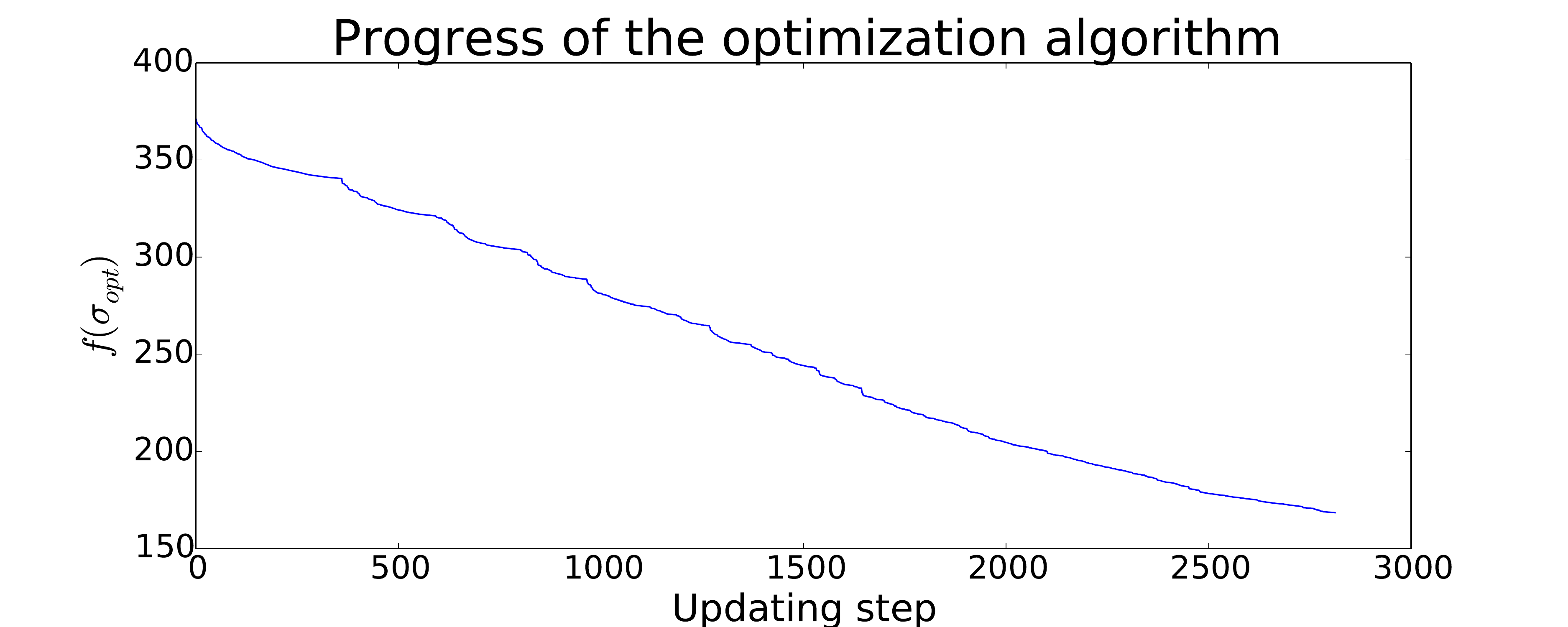}
\caption{Progress of the optimization algorithm for our data set. The initial value of the objective function is $370.99$ and 2,812 updating steps were performed. The final value of the objective function $f(\sigma_{\text{opt}})$ was equal to $168.5$.}
\label{fig:update_progress}
\end{figure}

The progress of the optimization algorithm is illustrated by means of Figure~\ref{fig:update_progress}.

The result of the algorithm will be the final assignment between atoms and bigrams defined by a permutation $\sigma_{\text{opt}} \in S_{2187}$ and the assignment rule $\alpha_{\sigma_{\text{opt}}(i)} \to \beta_i$.
This assignment is used to reconstruct the original bigram sets encrypted in the Bloom filters.

For example, the bigrams $\name{HE}_l$, $\name{E\textvisiblespace}_l$, $\name{\textvisiblespace K}_l$, $\name{RL}_l$, $\name{AR}_l$, $\name{LS}_l$, $\name{RU}_l$, $\name{KA}_l$, $\name{UH}_l$, $\name{SR}_l$, $\name{ER}_s$, $\name{R\textvisiblespace}_s$, $\name{CH}_s$, $\name{SC}_s$, $\name{HE}_s$, $\name{\textvisiblespace F}_s$, $\name{IS}_s$, $\name{ID}_s$, $\name{FI}_s$, $\name{N\textvisiblespace}_f$, $\name{\textvisiblespace S}_f$, $\name{ON}_f$, $\name{SI}_f$, $\name{IM}_f$ and $\name{MO}_f$
were assigned to the Bloom filter No. 850.

In the following section we describe how attribute values are reassembled from the reconstructed bigram sets.

\subsection{Reconstruction of Attribute Values}

In order to reconstruct the original attribute values of the records, we separated the bigrams belonging to different identifiers for each Bloom filter.
Then, our approach to reconstructing the original identifier values was to compare the obtained bigram sets with a list of bigram sets generated from reference lists of surnames, names and locations.
For Bloom filter No. 850, for example, an adversary would correctly guess that this Bloom filter encrypts a record belonging to the person \texttt{Simon Fischer} from the German city \texttt{Karlsruhe}.

\subsection{Results}

\noindent By using the approach described above, we were able to reconstruct 59.6\% of the forenames, 73.9\% of the surnames and 99.7\% of the locations correctly.
For 44\% of the $100,000$ records all the identifier values were recuperated successfully.

%-------- Section ------------

\section{Conclusion}\label{Conclusion}

\noindent
In this paper we demonstrate a successful fully automated attack on Bloom filters built from multiple identifiers.
We were able to recover approximately 77.7
\% of the original identifier values.
In contrast to the assumptions in \cite{Kuzu2012} and \cite{Niedermeyer2014}, that storing all identifiers in a single Bloom filter makes it more difficult to attack, we needed only moderate computational effort and publicly available lists of forenames, surnames, and locations to reconstruct the identifiers.
Note that there is no huge impact of the size of the database containing the Bloom filters. For our cryptanalysis it is sufficient to perform the attack on a subset of the given Bloom filters (100,000 as in our example should be adequate in most cases). Then for the remaining Bloom filters it would be sufficient to check for the atoms contained in those and to reconstruct the attribute values, since most assignments of atoms to bigrams are already known. Thus, the time needed for cryptanalysis is linear in the number of input Bloom filters. The time needed for the detection of atoms is $O(L^2)$ since there are $L$ possible values for the hash functions $f$ and $g$ in equation~\eqref{eqhash}.
Furthermore, the detection of atoms could easily be parallelized to make the computation faster and values of $L$ significantly larger than $L=1,000$ as considered in this paper would also have negative effects on the time needed for performing the linkage between two databases (note that in the large scale study reported in~\cite{Randall2014} a Bloom filter length of only $100$ was considered). Thus, the most time consuming step in our cryptanalysis should be the optimization algorithm presented in subsection \ref{subsec:correlation}.
Indeed, in the chosen parameter setup this procedure took about $402$ minutes on a notebook with $2.80$ GHz Intel$^\text{\textregistered}$ Core running Ubuntu 14.04 LTS.

To sum up, we do not recommend the usage of Bloom filters built from one or more identifiers, generated with the double hashing scheme, in applications where high security standards are required.
However, we applied our attack in a very special scenario, because the generated databases were encrypted using the double hashing scheme.
Thus, there are options for an improvement of the setting.

For example Niedermeyer et al. \cite{Niedermeyer2014} proposed several methods such as fake injections, salting or randomly selected hash values to harden the Bloom filters.
Hence, we are confident that methods like those proposed by Niedermeyer et al. show promise in the prevention of attacks like the one presented in this paper.

\section*{Acknowledgements}
This research has been partly supported by the grant SCHN 586/17-1 of the German Research Foundation (DFG) awarded to Rainer Schnell.

\newpage

\bibliographystyle{apalike}
{\small
\bibliography{references}}

\end{document}